\documentclass{article}
\usepackage[margin = 2cm]{geometry}
\usepackage{indentfirst}
\usepackage{graphicx}
\usepackage{float}
\title{Find The Optimized Structure of 5CBs}
\author{Zhi-Hao Chen and Jyh-Pin Chou}

\begin{document}
\maketitle
\begin{center}
Department of Physics, National Changhua University of Education, Changhua 50007, Taiwan
\end{center}

\section{Abstract}
    We use density functional theory (DFT) to investigate the geometric and electronic structures of multiple 5CB molecules. There are three parts for this research, test of the arrangement property of the 5CB molecule, calculation of two 4-n-pentyl-4’-cyanobiphenyl (5CB), and four molecules 5CB structure arrangement. First part reveals the result, which the two 5CBs pointing in opposite directions case possesses lower total energy, that has the same result of the optimized caused by the nematic property of 5CBs. In the result of two 5CBs calculation, two 5CBs optimized structures always appears at least one pair of parallel plains formed by the head benzene of one 5CB and the carbon chain. In the four 5CBs optimized structure case, we can find out several pairs of parallel plains consist of benzene plains and one parallel plain in two carbon chain.
\section{Introduction}
    Liquid crystal (LC) is state of matter that possesses the properties of solid crystals and the nature of liquids. LCs can be classified by the arrangement of their crystal structures, i.e., nematic, smectic, cholesteric, and discotic LCs, or by the procedures of synthesis, i.e., thermotropic and lyotropic LCs [1]. Within the classification, nematic LCs are often thought of as a state of matter that every layer of smectic LCs crash and all particles in this structure mix together, and the thermotropic LCs means that the LC phases occurs in a certain temperature range.\newline
    
    A classic double-benzene molecule in the field of LC is called 4-Cyano-4'-pentylbiphenyl (5CB), whose chemical structural formula is CH3(CH2)4C6H4C6H4CN, and 5CB molecules are nematic and thermotropic LCs. Because the 5CB molecules have an available nematic phase near room temperature and its relative simple structure, 5CB molecule is one of the most studied liquid crystal materials [2]–[7]. The arrangements of these molecules are in the same direction, which are usually defined as the guide shaft. And 5CB molecules are positive dielectric anisotropic molecules, such that 5CB molecules have polarity property. After applying the electric field on 5CBs, most of them have a parallel arrangement. The structure schematic diagram is shown in figure 1. Thus, the 5CBs are widely applied in the researches such as metamaterial [3], lenses [4], controllable diffraction gratings [5], etc.  . Because 5CB material is used in plenty of applications, these are also some researches discussing the properties about 5CB material, and most of them are carried out by the experiment, such as the crystalline property about 5CB [6], electro-optical and dielectric properties [7], etc  .
    \newline
    
    \begin{figure}[H]
    \centering
    \setlength{\abovecaptionskip}{0.cm}
    \setlength{\belowcaptionskip}{-0.cm}
    \includegraphics[width = 10 cm, height = 6 cm]{"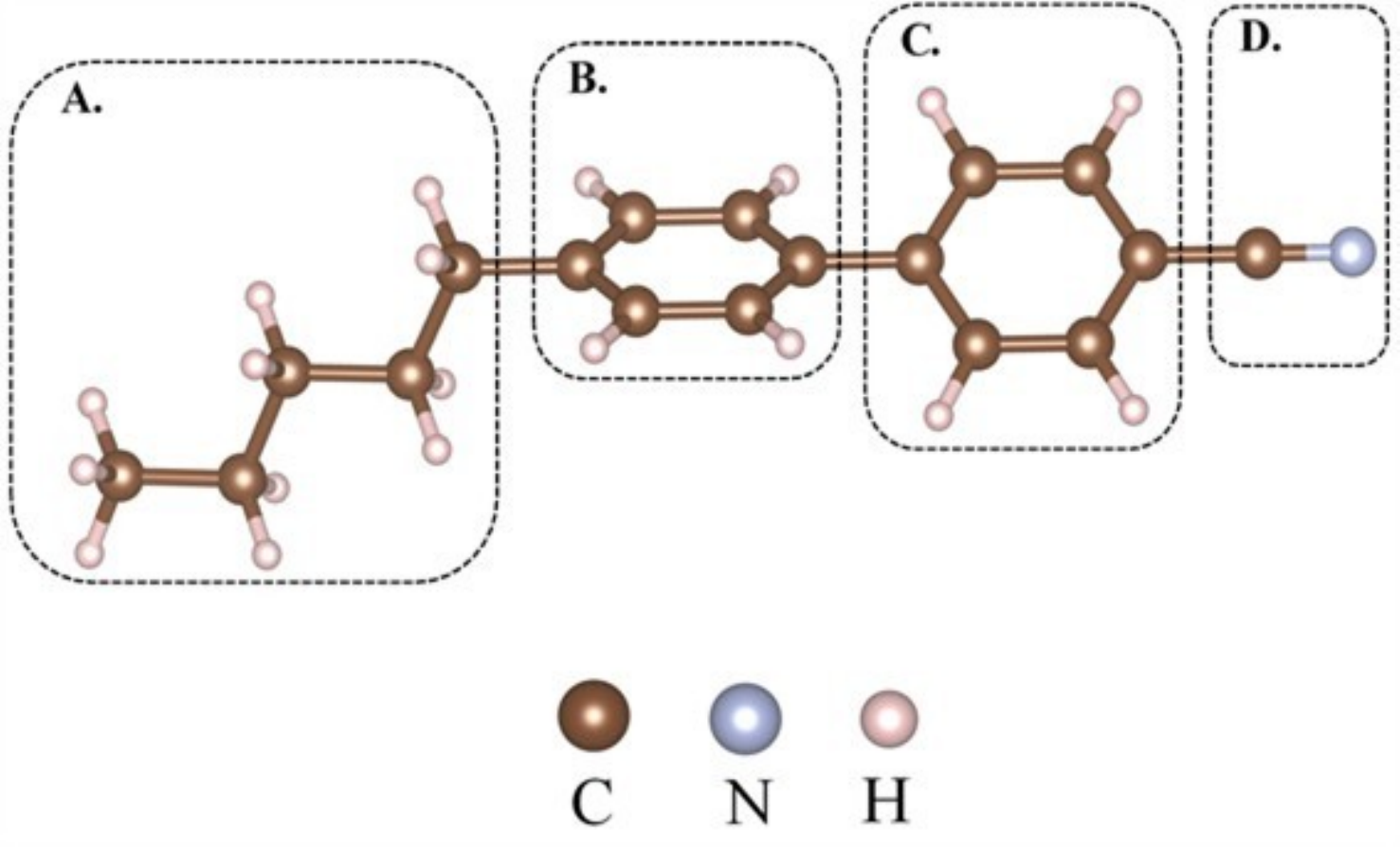"}
    \caption{We divide the 5CB molecule into four parts, A. carbon chain tail, B. benzene-tail, C. benzene-head, and D. head. The carbon, nitrogen, and hydrogen atoms are shown at the bottom of the figure.}
    \end{figure}
    
    Owing to the wide adoption of 5CB material in various researches, we want to discuss the geometric configuration of multiple 5CBs in 5CB material. Thus, in this article, we will build a series of well-defined multiple 5CB molecule structures to help is find the global minimum of the total system energy of 5CBs. It takes a long time of using VASP to calculate the enormous molecules with plenty of 5CB molecules, so we began this study from the structure of two 5CBs model and focused on finding the relations between the two 5CBs and four 5CBs optimized structure. If there is a strong relation correlation between two 5CBs and four 5CBs optimized structures by VASP, we could infer the optimized structure model of larger number of 5CB molecules model.
    
    \begin{figure}[H]
    \centering
    \setlength{\abovecaptionskip}{0.cm}
    \setlength{\belowcaptionskip}{-0.cm}
    \includegraphics[width = 13 cm, height = 10 cm]{"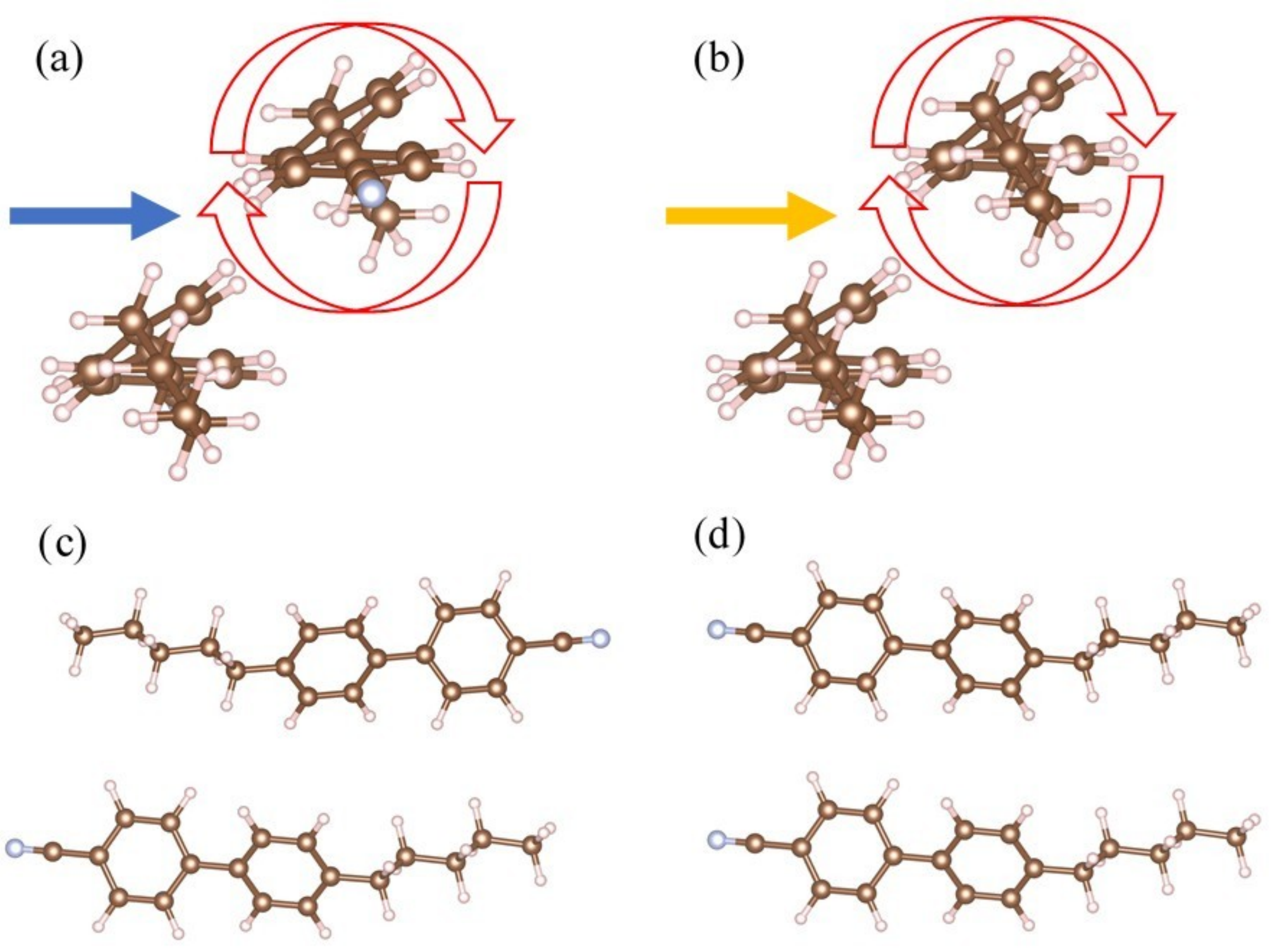"}
    \caption{(a) The red rotating-arrows show the relative rotating direction of two 5CBs in the opposite direction arrangement structure. (b) The red rotating-arrows show the relative rotating direction of two 5CBs in the same direction arrangement structure. (c) The image of looking into (a) through the blue arrow direction. (d) The image of looking into (b) through the yellow arrow direction.}
    \end{figure}
    
\section{Methodology}
    All calculations were performed by using Vienna Ab-Initio Simulation Package (VASP)[8], with projector-augmented wave (PAW)[9] pseudopotential. A plane wave cutoff energy of 450 eV was used in all calculations. The exchange-correlation functional was approximated by using Perdew-Burke-Ernzerhof within the generalized gradient approximation (PBE-GGA)[10]. And we considered Van Der Waal interaction with zero damping DFT-D3 method of Grimme[11]. \newline
    
    In the first part, we need to check out the arrangement properties of 5CBs. Before using the DFT calculation to optimize the structure of two 5CB molecule structure. We can build the two 5CBs structures in two different kinds of molecular arrangements, which are the opposite direction arrangement structure, figure 2(a) and figure 2(c), and the same direction arrangement structure, figure 2(b) and figure 2(d). Then, we use the total energy and Bader charge analysis to discuss which arrangement, figure 2(a) and 2(b), is corresponding to a more stable molecular system structure.     \newline
    
    Bader charge analysis is a method of identifying the areas occupied by each atom within the whole structure system  . The definition of atoms proposed by Richard Bader is based on the electronic charge density [12]. He used so-called zero point flux surfaces, where the charge density is a minimum perpendicular to the surface. Such that, the charge distribution in any region enclosed by the zero point flux surfaces is a well-defined approximation to the total charge of an atom [13]. We use the Bader charge analysis codes fromTransition State Tools for VASP (VTST)[14] to evaluate the electronic charge distribution. Electronic charge gain of an atom is defined as:
    \large
    $$\rho_g = \rho_B - \rho_v$$
    Where $\rho_g$, $\rho_B$ and $\rho_v$ are the electronic charge gain, Bader charge and valence electrons of the atom respectively.\newline
    
    In the second part, two 5CBs optimization, we can build a series of structures, which are based on the result of the test in the first part, by acting the relative rotation on the two 5CB molecules, same as the establishment of the structure in first part, but with more structures we built during the relative rotation.\newline
    
    Next, we want to build the four 5CBs structures. However, the degrees of freedom of the relative rotations in four 5CBs structure are too large to build, the number of structural possibilities is too large for us to evaluate. Such that we use Ab-Initio random structure searching (AIRSS) to help us build the structures of four 5CBs. With two and four 5CBs structure cases, we can compare the common characteristics between them.\newline
    
    AIRSS, what we used in the optimization of four 5CB molecule model, is an algorithm to find the optimized structure directly. First, AIRSS generates the positions of every atoms by random number method. Then, it reduces the total energy in a lattice until every net force acting on all atoms in a lattice is approaching to 0 and derives the most stable structure in the end. At the moment the potential energy surface has a local or global minimum. This method, which has been proved by experiments, is available for finding the most stable many kind of objective models[14].\newline
    
\section{Result and discussion}
    \subsection{Test of the arrangements of two 5CBs}
    
    In total energy test, we choose the distance between the two 5CBs about 9.05 Å, which prevents two molecules crash together during the relative rotation. And we built each structure per 15 degrees. Figure3 shows that the opposite direction arrangement structures are more stable than the same direction ones. The result of VASP simulation is consistent with the polarity property of the 5CB molecules.\newline
    \begin{figure}[H]
    \centering
    \setlength{\abovecaptionskip}{0.cm}
    \setlength{\belowcaptionskip}{-0.cm}
    \includegraphics[width = 11.5 cm, height = 7 cm]{"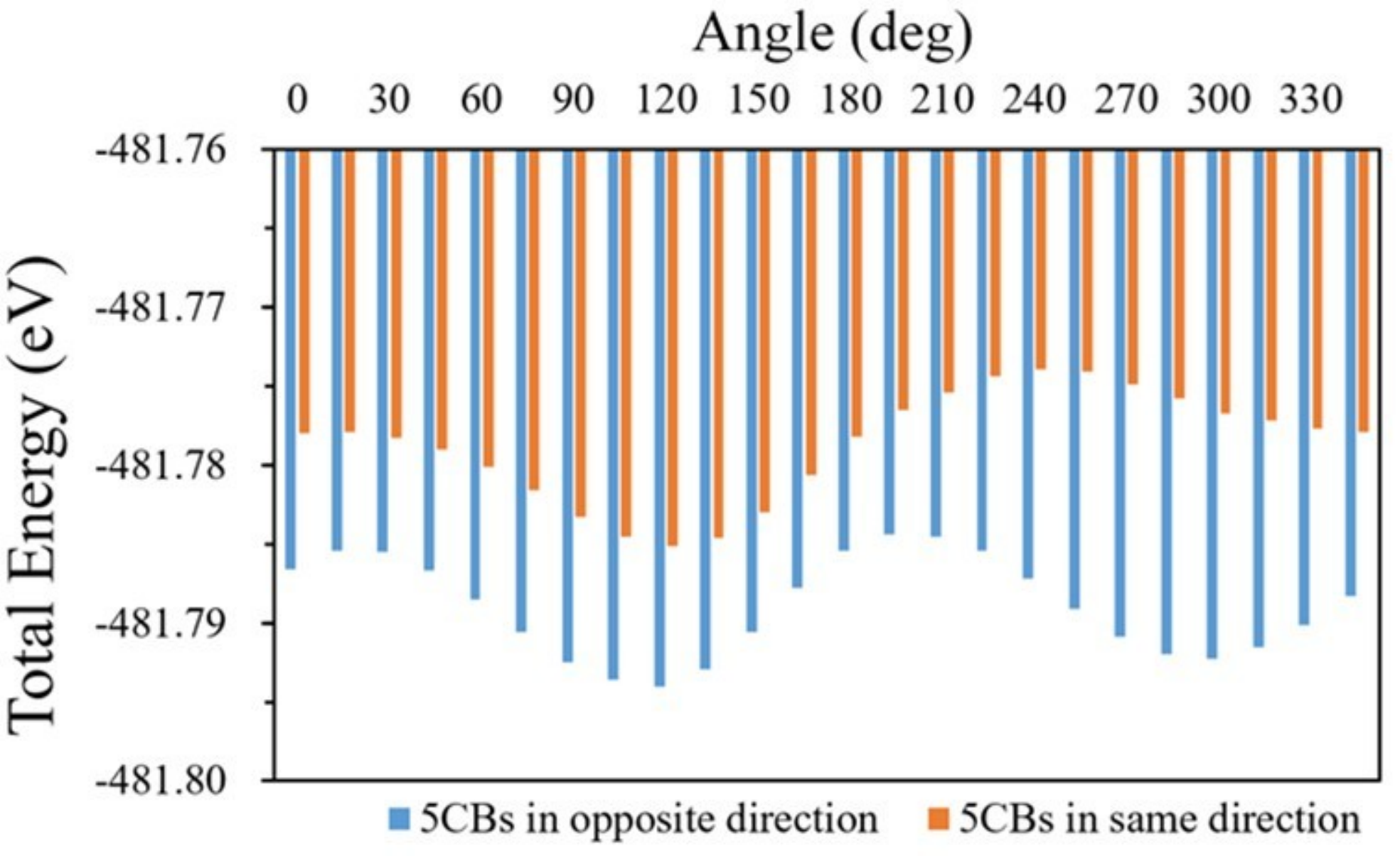"}
    \caption{The energy of the opposite direction arrangement is lower than the same direction ones.}
    \end{figure}
    
    \begin{figure}[H]
    \centering
    \setlength{\abovecaptionskip}{0.cm}
    \setlength{\belowcaptionskip}{-0.cm}
    \includegraphics[width = 18 cm, height = 5 cm]{"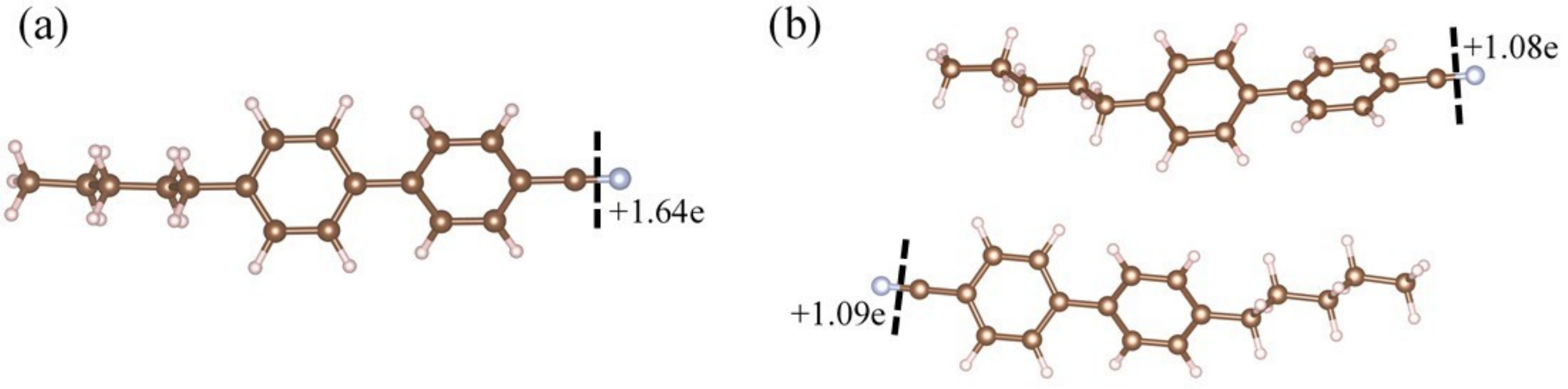"}
    \caption{The dashed-lines divide the nitrogen atoms and the rest parts of the 5CB molecules. (a) The value of charge gained by the nitrogen atom is 1.64 electronic charge. (b) The value of charge gained by the nitrogen atoms are 1.08 and 1.09 electronic charge respectively.}
    \end{figure}
    
    In the Bader charge analyzation of single 5CB and two 5CBs, as shown in figure 4(a) and 4(b), the properties of polarity are observed. The charge value gained by the nitrogen in figure 4(a) is larger than that in figure 4(b), so we could confirm that the polarity of 5CB molecule relates to the nitrogen atom and that the nematic properties of the 5CB molecule is available. The nematic properties of the 5CB molecule determine the opposite direction arrangement as a more stable model that has the same result of the total energy test. 
    
    \subsection{Two 5CBs structure}
    
    \begin{figure}[H]
    \centering
    \setlength{\abovecaptionskip}{0.cm}
    \setlength{\belowcaptionskip}{-0.cm}
    \includegraphics[width = 12 cm, height = 6 cm]{"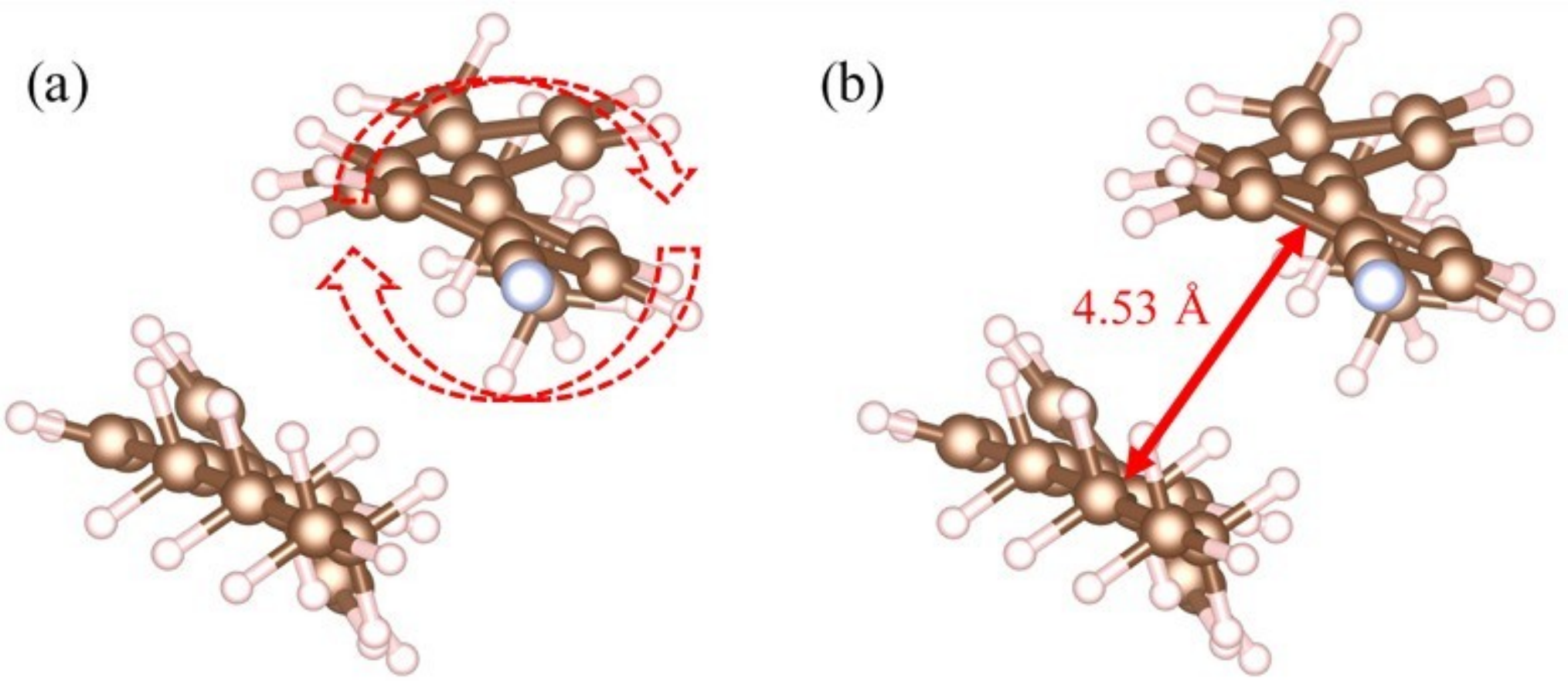"}
    \caption{(a) We fixed one 5CB molecule and rotated the other one. The direction of rotation is along the axis pointing outward from the figure. (b) The distance between the two 5CB molecules that we set is about 4.53 Å.}
    \end{figure}
    
    After the test, we use the opposite direction arrangement to find the optimized structure of two 5CBs case. Besides, we let the distance between the two 5CBs is about 4.53 Å. Then we change the relative rotation angle of the two 5CBs by rotating one of 5CBs and fix the other one, shown in figure 5, and the result is shown in figure 7.\newline
    
    In two 5CBs structure result, see figure 6, after DFT calculation, the data points with relative total energies lower than 0.03eV are colored by red, in region B, and the data points with the energies beyond 0.125eV are colored by yellow, in region A. The energy differences between the structures in region A and region B are larger than the thermal agitation in room temperature, which is about 20~30 meV. Thermal agitation is kBT [15], which kB is Boltzmann constant, and T is temperature under Kelvin scale. Such that, there are some significant most stable structures that we can find even with the existence of thermal agitation energy.\newline
    
    The optimized structure with 0-degree relative rotation angle structure, one of the structures in region B, in figure 6 is shown in figure 7. we can find that a pair of parallel plains which is comprised of a carbon chain and a head benzene. Also, we can always find out the parallel plains in the other structures in region B. In contrast to the structures inside the region B, the structures in region A don’t always have the parallel property between the carbon chain and head benzene. The included angle between the head benzene and the carbon chain, schematic diagram of it is in figure 7, for all structures result in figure 6 is shown in figure 8. In figure 8, the yellow and red dots is corresponding to the structures in region A and region B respectively in figure 6, where the red points correspond to more stable structures. All the included angle of red point data are smaller than 10 degrees, which means the parallel relation between the benzene and carbon chain is a significant characteristic of the higher stability structure.\newline
    
    \begin{figure}[H]
    \centering
    \setlength{\abovecaptionskip}{0.cm}
    \setlength{\belowcaptionskip}{-0.cm}
    \includegraphics[width = 14.5 cm, height = 9 cm]{"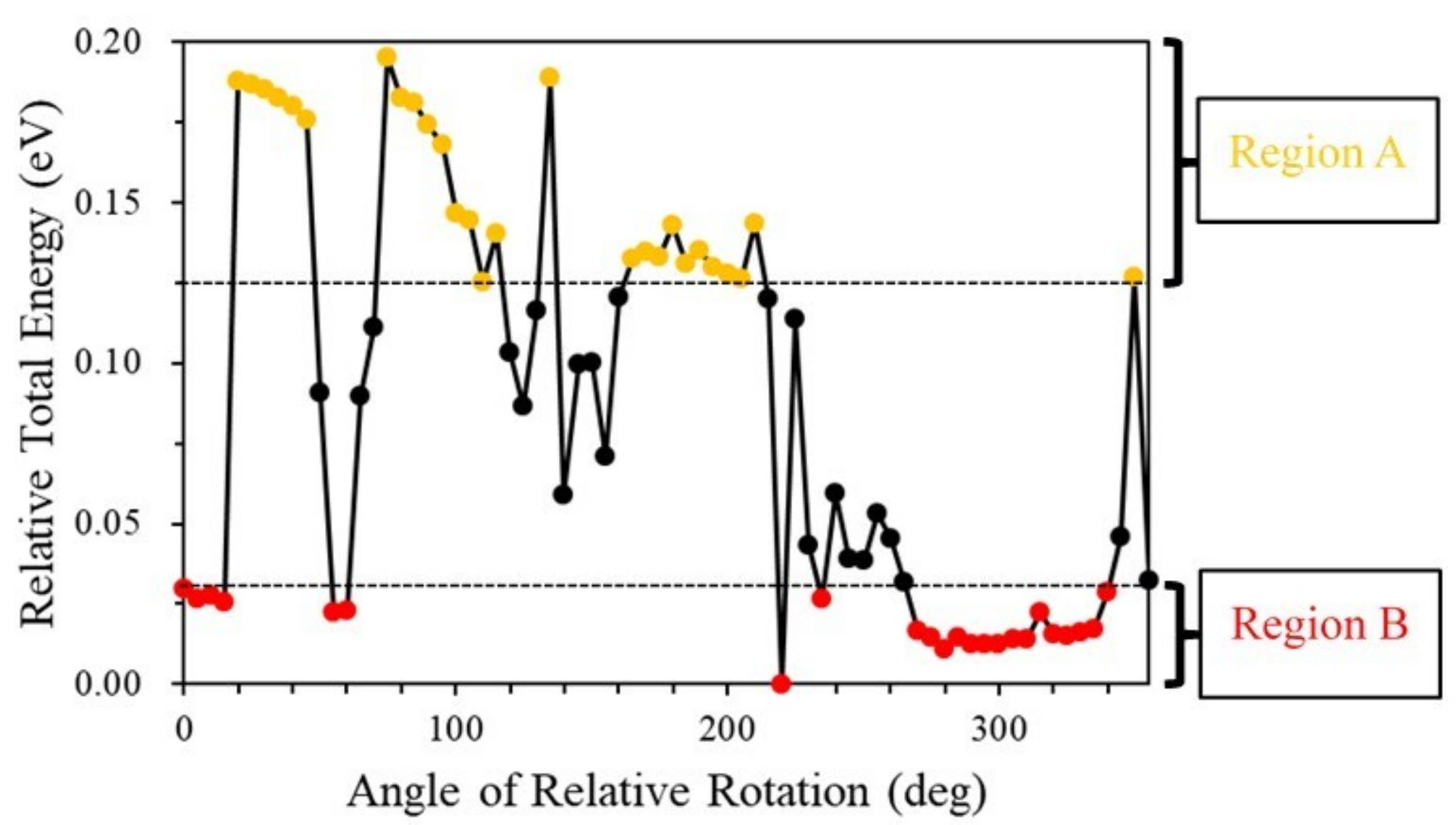"}
    \caption{We built each structure per 5 degrees, then we generated 72 structures in all. The red dots correspond to the optimized structures with the relative total energies lower than 0.03eV, in region B, and the yellow dots correspond to the structure with the relative energies higher than 0.125eV, region A.}
    \end{figure}
    
    \begin{figure}[H]
    \centering
    \setlength{\abovecaptionskip}{0.cm}
    \setlength{\belowcaptionskip}{-0.cm}
    \includegraphics[width = 13 cm, height = 8 cm]{"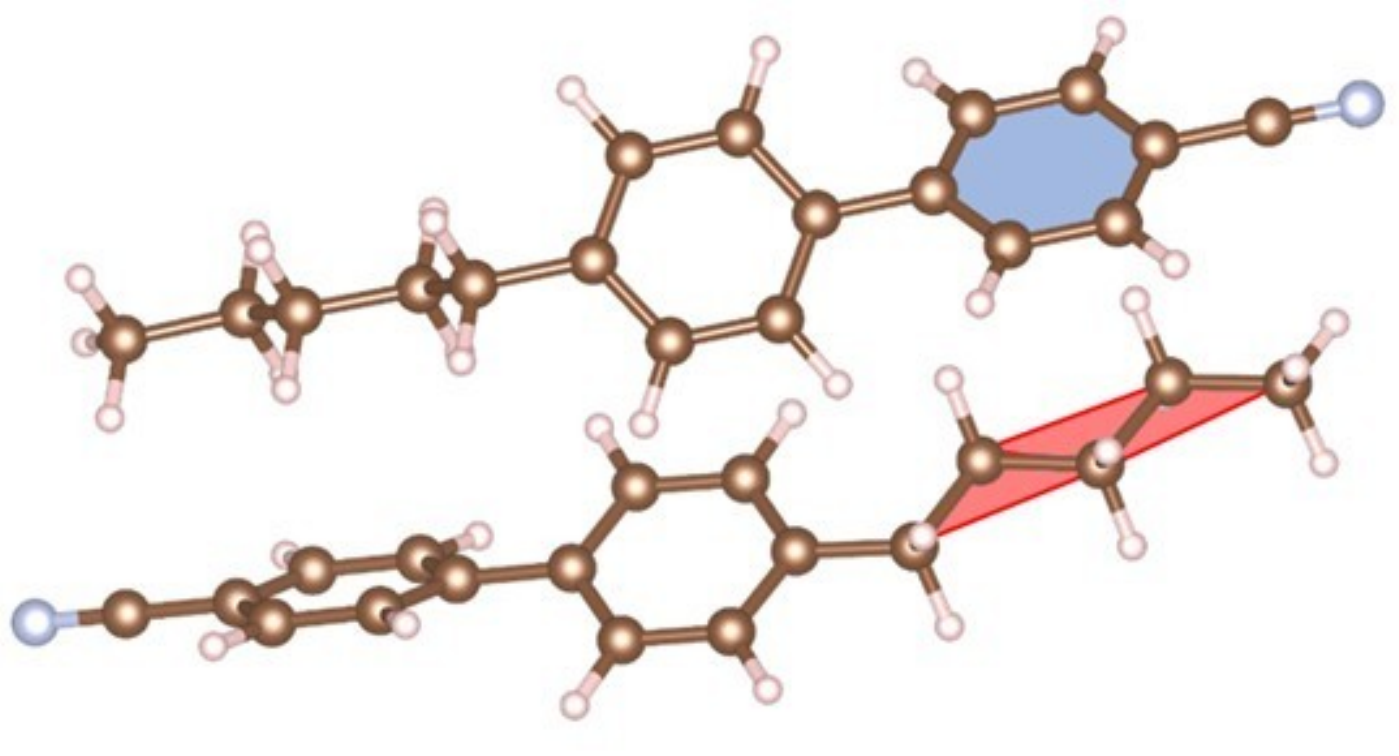"}
    \caption{Optimized structure with the angle of relative rotation is 0 in figure 6. In this structure result, the colored region is the parallels, which is composed of a benzene (blue region) of one 5CB and a carbon chain (red region) of another 5CB. }
    \end{figure}
    
    \begin{figure}[H]
    \centering
    \setlength{\abovecaptionskip}{0.cm}
    \setlength{\belowcaptionskip}{-0.cm}
    \includegraphics[width = 13 cm, height = 8 cm]{"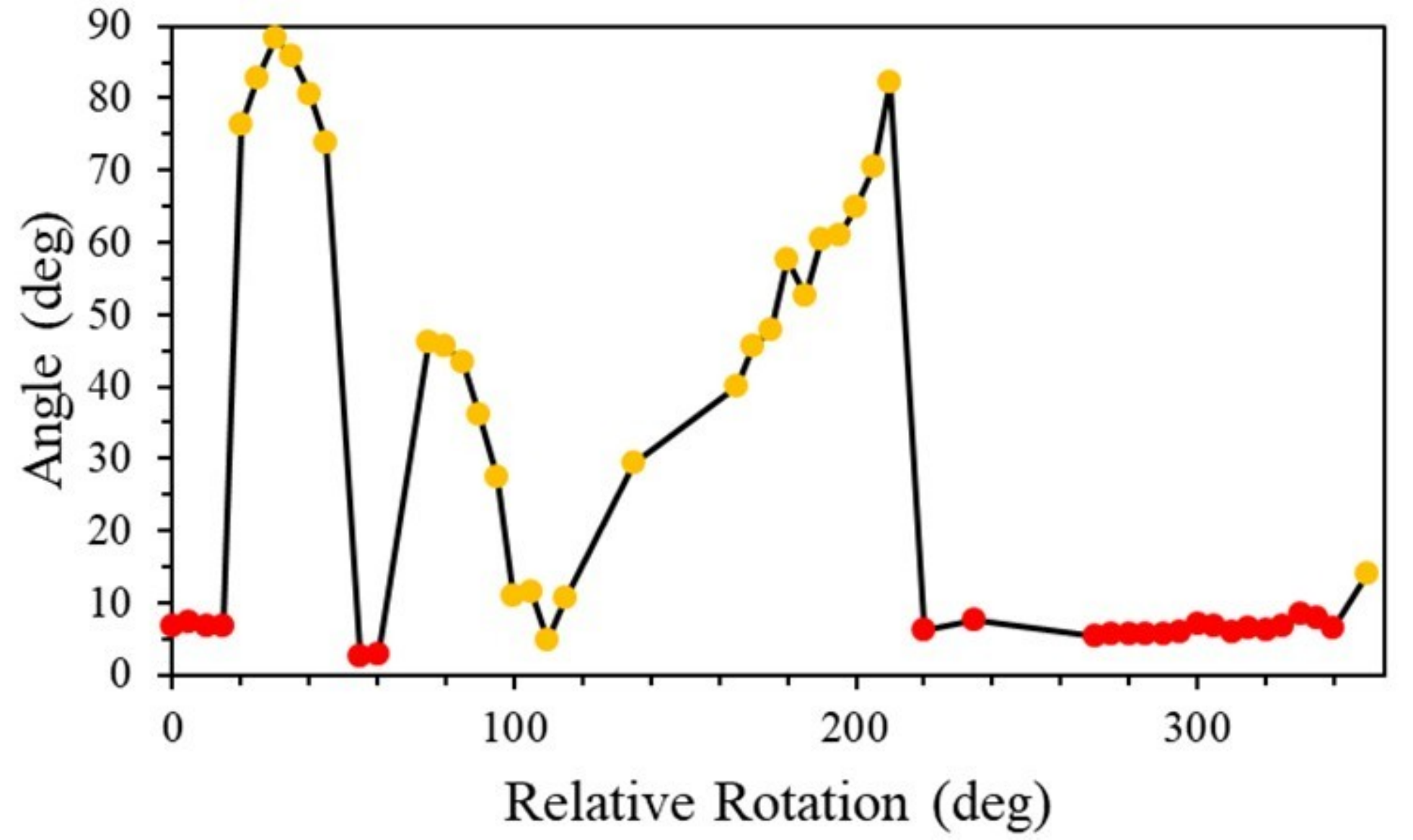"}
    \caption{The red and yellow dots are corresponding to the colored points in figure 6. In the optimized structures of the red colored points, every included angle between the head benzene and carbon chain, the two plains shown in figure 7, is always smaller than 10 degrees. Different from the red colored data, the yellow ones are corresponding to larger included angle between the two plains with respect to the red colored data.}
    \end{figure}
    
    \subsection{Four 5CBs structure}
    Because all structures built by AIRSS are random, we let all optimized structure data points, 500 data points in total, which are shown in the figure 9(a), follow the order of relative total energy to sequence. In figure 9, we only took 20 lowest total relative energy data points. We could notice that in the red and green circle exist larger gaps, which means the yellow data points have similar optimized structure, so as brown data points. After the analyzation of the relative total energy sorting diagram, we want to analyze the structure characteristics of yellow and brown points.\newline
    
    \begin{figure}[H]
    \centering
    \setlength{\abovecaptionskip}{0.cm}
    \setlength{\belowcaptionskip}{-0.cm}
    \includegraphics[width = 17 cm, height = 6 cm]{"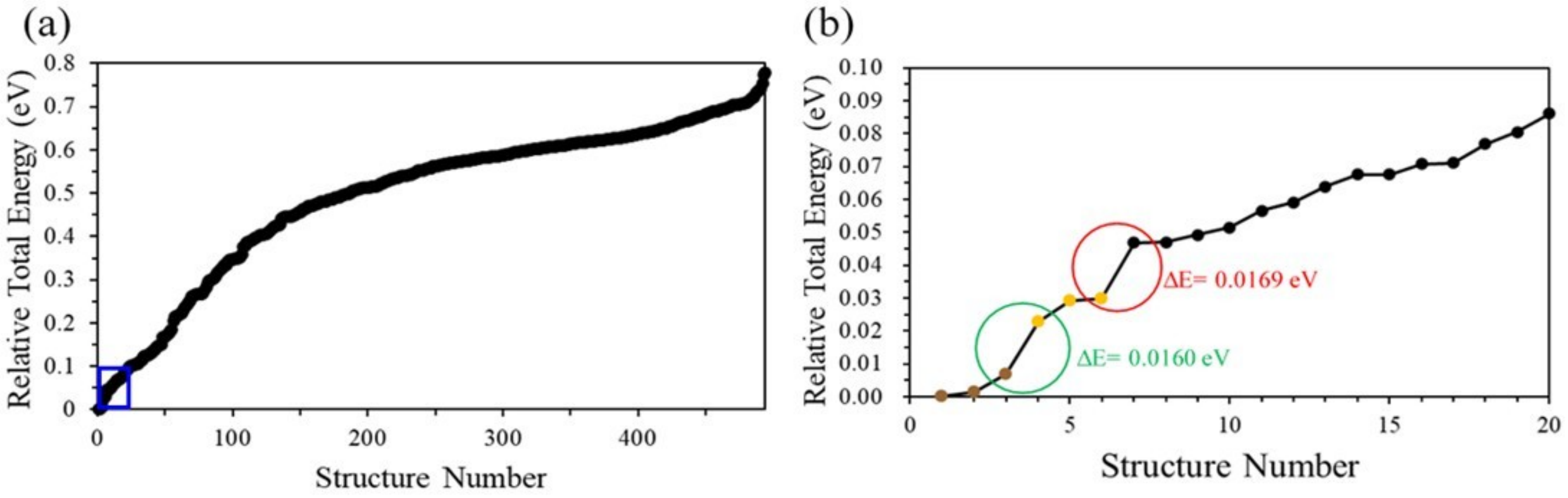"}
    \caption{(a) There are 494 structure data in the curve, and the structure numbers are following the order of the energies, from the lowest to the highest. (b) Partially highlighted region, enclosed by the blue square in (a), with 20 lowest relative energies of all structure data. The brown dots are the structures with closed relative total energies, and so as yellow ones. There are two larger energy gaps enclosed by red and green circles, and their corresponding values are 0.0169eV and 0.0160eV.}
    \end{figure}
    
    Within these data, we take the lowest energy data point structure, for example. In figure 10(a), A, B, C, and D are four 5CB molecules. The benzene of head molecule A is almost parallel to the head benzene of C, and the tail benzene of A is almost parallel to the tail benzene of C (Figure 10(d)). Also, carbon chain of A is almost parallel to carbon chain of C (Figure 10(c)). Same as A and C case, B and D molecules also have these relations. The result of Four 5CB molecules is different to the result of two 5CB molecules case because we can’t find the parallel relation between a benzene of one molecule and a carbon chain of another molecule.\newline
    
    \begin{figure}[H]
    \centering
    \setlength{\abovecaptionskip}{0.cm}
    \setlength{\belowcaptionskip}{-0.cm}
    \includegraphics[width = 12 cm, height = 11 cm]{"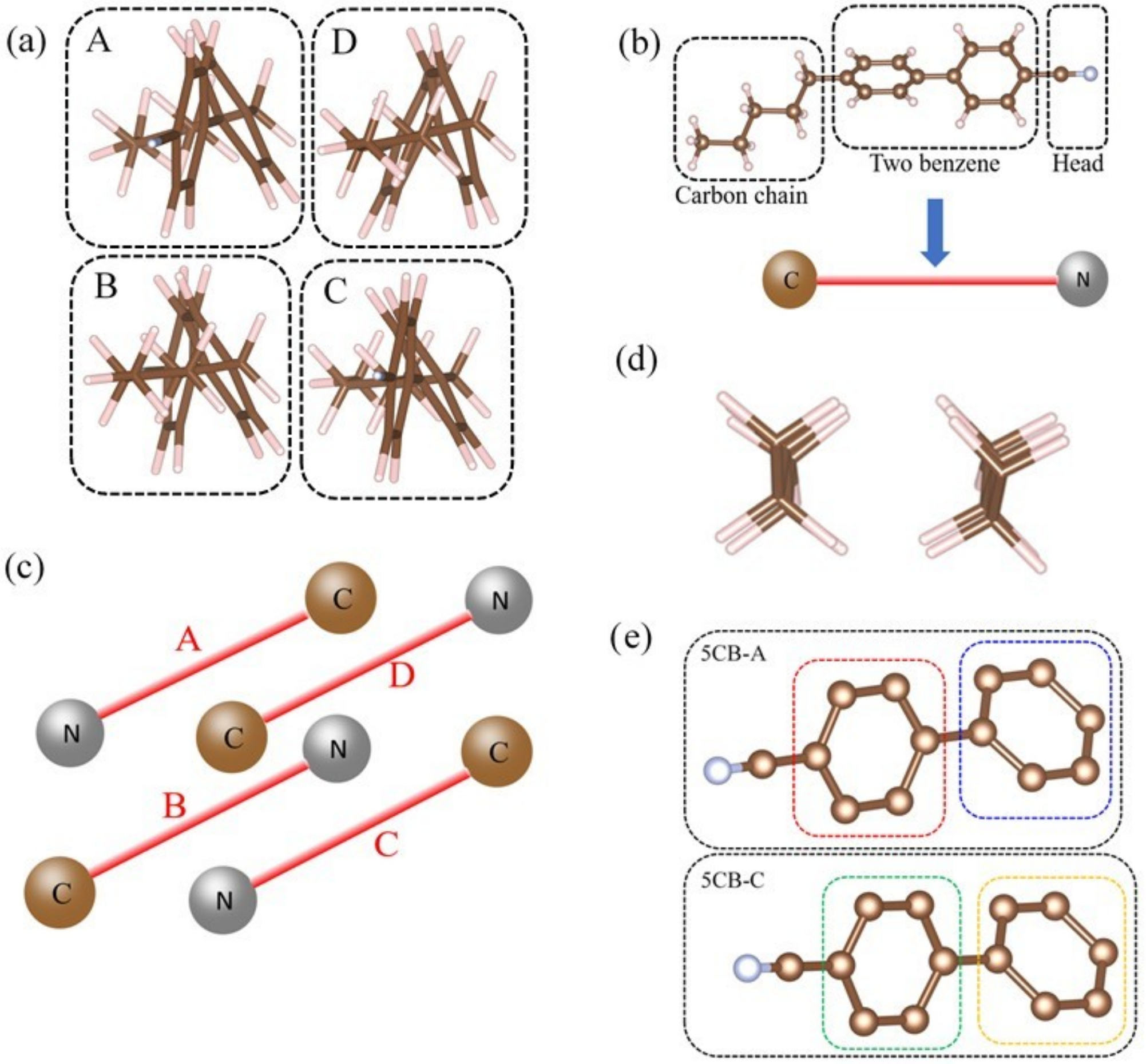"}
    \caption{(a) We divided the four 5CBs structure into four parts, A, B, C, and D. Each of them is one molecule of the four 5CBs structure. (b) We divide the 5CB molecule into carbon chain, two benzene, and head three parts, and replace the 5CB molecule by the ball-stick model, which is composed of the brown sphere with alphabet C, the red stick, and the gray sphere with alphabet N, and they are corresponding to the carbon chain, two benzene, and head respectively. (c) The simplified figure of (a), and these four molecules are corresponding to A, B, C, and D. (d) Carbon chains of A and C are almost parallel to each other. The included angle between these carbon chains is about 7.71 degrees. (e) The benzenes of A and C are almost parallel to each other. The included angle between head benzene of A (in red dashed square) and the head benzene of C (in green dashed square) is about 4.00 degrees. The included angle between tail benzene of A (in blue dashed square) and the tail benzene of C (in yellow dashed square) is about 3.14 degrees.}
    \end{figure}
    
\section{Conclusion}

    Three parts of this study focus on analyzing the geometry structure and the corresponding relative total energy, and the optimized structure possess the minimum energy of all structure.\newline
    
    The first part of them tells us that the stable arrangement of two 5CB molecules under the calculation executed by DFT-D3 calculation with 0K temperature reveals the nematic property of 5CB molecule, which determines the basic framework of the structure used in others two parts.\newline
    
    In the second part, two 5CBs structure optimization, the parallel plains between the head benzene of one molecule and the carbon chain of the other one can’t be ignored, because the structures with lower total energy possess this special geometry characteristic. And there are several energy gaps larger than 0.1eV, which means the phenomenon of optimization isn’t easy to be overwhelmed by the thermal agitation energy in room temperature, about 300K.\newline
    
    Final part is the optimization of structure with four 5CBs. In this case, benzene-benzene and chain-chain parallel plains are the main characteristics of optimization, and there doesn’t exist parallel plains with the combination of carbon chain and head benzene that we found in two 5CB molecules structure. Thus, we know that we can’t directly build a structure very similar to optimized four 5CBs structure by combining the optimized result of two 5CBs case. The energy gaps have only 16 ~ 17 meV, which is smaller than the thermal agitation energy under the room temperature, that means the four 5CBs structures have no strong tendency of geometry characteristic in the optimization process.\newline
    
    From two 5CBs to four 5CBs, we find that the benzene ring isn’t always dominating the tendency of optimization, and we can’t describe the optimized structure of four 5CBs by two 5CBs case. Such that, an optimized structure, in which the 5CBs are stacked in the same arrangement we set in four 5CBs case, with more than four 5CB molecules is quite hard to discuss by the simpler structure with less 5CBs in it, and the arrangement of large number 5CBs tends to be affected by room temperature instead in orderliness. However, the optimization tendency of the structure with only two 5CB molecules is still obvious in room temperature, this result may be useful in the two-dimension 5CBs arrangement optimization.\newline
    
\section{Reference}
\noindent 
[1]	H. Kelker, History of Liquid Crystals, Molecular Crystals and Liquid Crystals, vol. 21, no. 1–2, pp. 1–48, Jan. 1973, doi: 10.1080/15421407308083312.\newline
\noindent
[2]	P. Selvaraj, K. Subramani, B. Srinivasan, C.-J. Hsu, and C.-Y. Huang, Electro-optical effects of organic N-benzyl-2-methyl-4-nitroaniline dispersion in nematic liquid crystals, Sci Rep, vol. 10, no. 1, p. 14273, Dec. 2020, doi: 10.1038/s41598-020-71306-1.\newline
\noindent
[3]	D. Shrekenhamer, W.-C. Chen, and W. J. Padilla, Liquid Crystal Tunable Metamaterial Absorber, Phys. Rev. Lett., vol. 110, no. 17, p. 177403, Apr. 2013, doi: 10.1103/PhysRevLett.110.177403.\newline
\noindent
[4]	H. E. Milton, H. F. Gleeson, P. B. Morgan, J. W. Goodby, S. Cowling, and J. H. Clamp, Switchable liquid crystal contact lenses: dynamic vision for the ageing eye, San Francisco, California, United States, Feb. 2014, p. 90040H. doi: 10.1117/12.2044676.\newline
\noindent
[5]	J. Zhang, V. Ostroverkhov, K. D. Singer, V. Reshetnyak, and Yu. Reznikov, Electrically controlled surface diffraction gratings in nematic liquid crystals, Opt. Lett., vol. 25, no. 6, p. 414, Mar. 2000, doi: 10.1364/OL.25.000414.\newline
\noindent
[6]	T. Hanemann, W. Haase, I. Svoboda, and H. Fuess, Crystal structure of 4-pentyl-4-cyanobiphenyl (5CB), Liquid Crystals, vol. 19, no. 5, pp. 699–702, Nov. 1995, doi: 10.1080/02678299508031086.\newline
\noindent
[7]	H. Suzuki, A. Inaba, J. Krawczyk, and M. Massalska-Arodź, Thermodynamic properties of chiral liquid crystalline material (S)-4-(2-methylbutyl)-4-cyanobiphenyl (5CB), The Journal of Chemical Thermodynamics, vol. 40, no. 8, pp. 1232–1242, Aug. 2008, doi: 10.1016/j.jct.2008.04.001.\newline
\noindent
[8]	G. Kresse and J. Furthmüller, Efficient iterative schemes for ab initio total-energy calculations using a plane-wave basis set, Phys. Rev. B, vol. 54, no. 16, pp. 11169–11186, Oct. 1996, doi: 10.1103/PhysRevB.54.11169.\newline
\noindent
[9]	P. E. Blöchl, Projector augmented-wave method, Phys. Rev. B, vol. 50, no. 24, pp. 17953–17979, Dec. 1994, doi: 10.1103/PhysRevB.50.17953.\newline
\noindent
[10]	J. P. Perdew, K. Burke, and M. Ernzerhof, Generalized Gradient Approximation Made Simple, Phys. Rev. Lett., vol. 77, no. 18, pp. 3865–3868, Oct. 1996, doi: 10.1103/PhysRevLett.77.3865.\newline
\noindent
[11]	S. Grimme, J. Antony, S. Ehrlich, and H. Krieg, A consistent and accurate ab initio parametrization of density functional dispersion correction (DFT-D) for the 94 elements H-Pu, The Journal of Chemical Physics, vol. 132, no. 15, p. 154104, Apr. 2010, doi: 10.1063/1.3382344.\newline
\noindent
[12]R. F. W. Bader, Atoms in molecules, Acc. Chem. Res., vol. 18, no. 1, pp. 9–15, Jan. 1985, doi: 10.1021/ar00109a003.\newline
\noindent
[13]Henkelman Research Group,http://theory.cm.utexas.edu/henkelman/code/bader/ (accessed Sep. 12, 2021).\newline
\noindent
[14] Transition State Tools for VASP, https://theory.cm.utexas.edu/vtsttools/ (accessed Sep. 12, 2021).\newline
\noindent
[15] C. J. Pickard and R. J. Needs, Ab initio random structure searching, J. Phys.: Condens. Matter, vol. 23, no. 5, p. 053201, Feb. 2011, doi:10.1088/0953-8984/23/5/053201.\newline
\noindent
[16] Atkins’ Physical Chemistry - Paperback - Peter Atkins, Julio de Paula, James Keeler - Oxford University Press. (accessed Jun. 28, 2021).

\end{document}